\documentclass[10pt, conference]{ieeeconf}      

\IEEEoverridecommandlockouts                              
\usepackage{graphicx}          

\usepackage{epstopdf}
\usepackage{times,mathptmx}
\usepackage[T1]{fontenc}
\usepackage{color}
\usepackage{pstool}
\usepackage[latin1]{inputenc}
\usepackage{rotating}
\usepackage{amsmath}
\usepackage{amssymb}
\usepackage{graphicx}
\usepackage{psfrag}
\usepackage{subfigure}
\usepackage{macros}
\usepackage{booktabs}
\usepackage{tikz}
\usepackage{url}
\usetikzlibrary{shapes,arrows}
\renewcommand{\baselinestretch}{0.973}

\title{\LARGE \bf Application Set Approximation in Optimal Input Design for\\ Model Predictive Control}

\author{Afrooz Ebadat*, Mariette Annergren*, Christian A. Larsson*, Cristian R. Rojas*, and Bo Wahlberg*
\thanks{This work was partially supported by the Swedish Research Council and the Linnaeus Center ACCESS at KTH, and the European Union's Seventh Framework Programme (FP7/2007-2013)}
\thanks{*Automatic Control Lab and ACCESS, School of Electrical Engineering, KTH, SE-100 44 Stockholm, Sweden.
 (e-mail: \{ebadat, marann, christian.larsson, crro, bo\}@ee.kth.se)}}

\begin{document}

\maketitle
\thispagestyle{empty}
\pagestyle{empty}

\begin{abstract}
This contribution considers one central aspect of experiment design in system identification. When a control design is based on an estimated model, the achievable performance is related to the quality of the estimate. The degradation in control performance due to errors in the estimated model is measured by an application cost function. In order to use an optimization based input design method, a convex approximation of the set of models that satisfies the control specification is required. The standard approach is to use a quadratic approximation of the application cost function, where the main computational effort is to find the corresponding Hessian matrix. Our main contribution is an alternative approach for this problem, which uses the structure of the underlying optimal control problem to considerably reduce the computations needed to find the application set. This technique allows the use of applications oriented input design for MPC on much more complex plants. The approach is numerically evaluated on a distillation control problem.
\end{abstract}

\section{Introduction}
System identification for control concerns the problem of using experimental data from a dynamical system to identify a model to be used for control design, see e.g.~\cite{1920,Forssell&Ljung:00,Lindqvist&Hjalmarsson:00,Belforte:02,
Jauberthie2006881,Rojas&al:07Aut,Rojas&Aguero&al:08aut,Franceschini20084846,Pronzato:08,IR-EE-RT_2007:003}.
The opportunity to also design the excitation input signal to be used in the experiment opens up for the possibility to connect the system identification experimental conditions to the required the control performance. One way is to formulate this as a convex optimization problem,\cite{Hildebrand:03,Jansson:04a,IR-EE-RT_2006:044,Geversetal:06a,Hjalmarsson&Jansson:06,Gerencser:07}.
We will study one aspect of the so-called applications oriented input design introduced in \cite{Hjalmarsson:2009}, specifically for model predictive control (MPC). The objective is to guarantee, with a given probability, that the estimated model belongs to the set of models that satisfies the control specifications. This objective can be stated mathematically as a set constraint where the set of all identified models corresponding to a particular level of confidence must lie inside the set of all models fulfilling the control specifications \cite{Hjalmarsson:2009}. To ensure that the obtained optimization problem is convex, we generally must make a convex approximation of the set constraint. Two known approaches of doing this are the scenario approach, \cite{Larsson2011} and \cite{Calafiore2006}, and the ellipsoidal approach, \cite{BoWahlbergHakanHjalmarson2010}. The main drawback of these methods are the computational efforts necessary to obtain a descent approximation. Both methods require several simulations to be made of the closed loop system with MPC. In this paper, we introduce a new method of approximating the set constraint with a convex one. The method is based on a perturbation analysis which only requires one simulation of the closed loop system. Thus, the method is expected to be much faster than both the scenario and the ellipsoidal approach.
The outline of the paper is as follows. In Section~\ref{Preliminaries}, we go-through the mathematical background necessary. We describe the scenario approach and the ellipsoidal approach in Section~\ref{Application set approximation}, followed by a detailed description of the proposed new method in Section~\ref{proposed method}. In Section~\ref{Numerical examples}, we illustrate the method in two numerical examples and in Section~\ref{Conclusions}, some conclusions are stated.
\section{Preliminaries}
\label{Preliminaries}
\subsection{System and model}
We consider a linear, time-invariant, asymptotically stable system in discrete time. The system is
\begin{equation}
\label{system}
\begin{split}
& x(k+1)=Ax(k)+Bu(k),
\\
& y(k)=Cx(k).
\end{split}
\end{equation}
Here, $x \in \R^{n_x}$ is the state vector, $u \in \R^{n_u}$ is the input vector and $y(k)\in\R^{n_y}$ is the output vector. The matrices $A$, $B$ and $C$ are the state space matrices of the system. In system identification, we want to find a model of the system \eqref{system}. We assume that the model is parametrized with an unknown parameter vector $\theta \in \R^n$, that is,
\begin{equation}
\label{model}
\begin{split}
& x(k+1, \theta)=A(\theta)x(k, \theta)+B(\theta)u(k, \theta),
\\
& y(k, \theta)=C(\theta)x(k, \theta).
\end{split}
\end{equation}
In addition, we assume that the model \eqref{model} matches system \eqref{system} exactly when $\theta=\thetao$. We call $\thetao$ the true parameter vector. The objective of system identification then is to estimate the values of $\theta$ that best describes the system, according to some measure. The estimated parameter vector, given $N$ measurements in the identification experiment, is denoted $\hat{\theta}_N$.
\subsection{Model predictive control}
Model predictive control (MPC), also referred as receding horizon control, is an advanced optimization based control technique. At each control interval, MPC computes a sequence of optimal inputs by solving an on-line optimization problem, where a model is used to predict the behavior of the plant. However, only the first input value is applied to the plant. A common optimization problem that is solved at every time instant $t$, in MPC is
\begin{equation}
\begin{split}
\label{MPCOID}
\displaystyle \min_{\substack{\{u(k,\theta)\}_{k=1}^{N_u} }}  &J\hspace{-0.1cm}=\hspace{-0.1cm}\sum^{N_y}_{k=0}\left\|y(k+1,\theta)-r(k+1)\right\|^{2}_{Q}+\sum^{N_u}_{k=1}\left\|\Delta u(k,\theta)\right\|^{2}_{R} \hspace{-0.3cm}
\\\hspace{-0.1cm}
\text{s. t. } 
& x(k+1,\theta)=A(\theta)x(k,\theta)+B(\theta)u(k,\theta),\hspace{0cm}k=1,...,N_y,
\\
& y(k+1,\theta)=C(\theta)x(k+1,\theta), \hspace{0.1cm}k=0,...,N_y,
\\
& x(1,\theta)=x^{*}(t,\theta),
\\
& \Delta u(1,\theta)=u(1,\theta)-u^{*}(t-1,\theta),
\\
& u_{min} \leq{u(k,\theta)}\leq{u_{max}},k=1,...,N_u,\\
& y_{\mathop{min}} \leq{y(k+1,\theta)}\leq{y_{max}},k=0,...,N_y.
\end{split}
\end{equation}
Here $r(k)$ is the reference trajectory, $\theta\in\R^n$ is the vector of system parameters, $Q$ and $R$ are weight matrices, $N_u$ and $N_y$ are control and prediction horizons, respectively, and $\Delta u(k,\theta)=u(k,\theta)-u(k-1,\theta)$. Note that $\Delta u(k,\theta) = 0$ for $k>N_u$. $u^{*}(t-1,\theta)$ is the optimal input value applied to the system at time instant $t-1$, and $x^{*}(t,\theta)$ is the system state at time $t$, which can be obtained by direct measurement or an observer.  Different MPC formulations are discussed in more detail in \cite{J.M.Maciejowski2002}.

\subsection{Prediction error method}
We use the prediction error method (PEM) to estimate the unknown parameters of a considered system. The unknown parameters are denoted $\theta\in\R^n$, the true parameters representing the system are denoted $\thetao\in\R^n$ and the estimated parameters given $N$ measurements are denoted $\hat{\theta}_N\in\R^n$. A key asymptotic ($N\rightarrow \infty$) property of PEM, is that the estimated parameters lie in an \emph{identification set} with a certain probability. The identification set is defined as  
\begin{equation}
\label {id set}
\Esi(\alpha)=\left\{\theta:[\theta-\thetao]^TI_F(\thetao)[\theta-\thetao]\leq  \frac{\chi^2_\alpha(n)}{N} \right\},
\end{equation}
where the term $\chi^2_\alpha(n)$ is the $\alpha$-percentile of the $\chi^2$-distribution with $n$ degrees of freedom and $I_F$ is the Fisher information matrix. We thus have that $\hat{\theta}_N\in\Esi(\alpha)$ with probability $\alpha$ when $N\rightarrow \infty$. For more details, we refer the reader to \cite{L.Ljung1999}.
\subsection{Applications oriented input design}
Model-based controllers, such as MPC, use a model in order to control a system. Therefore, the control performance is affected by any plant-model mismatch. We use the concept of an application cost function to relate the plant-model mismatch to the performance degradation. We use a scalar function of $\theta$ as the application cost and denote it $V_{app}(\theta)$. The cost function is chosen such that its minimum value occurs at $\theta =\thetao$. In particular, we assume that $V_{app}(\theta_0) = 0$. Note that if $\Vapp(\theta)$ is twice differentiable in a neighborhood of $\theta_0$, this implies that 
\begin{equation}
\Vapp(\thetao) = 0 \hspace{0.1cm} , \hspace{0.1cm} \Vapp'(\thetao) = 0 \hspace{0.1cm} \mathrm{and} \hspace{0.1cm}\Vapp''(\thetao) \succeq 0,
\nonumber
\end{equation}  
see \cite{BoWahlbergHakanHjalmarson2010}. For a given plant, there is a limit on the maximum value of acceptable performance degradation, that is,
\begin{equation}
\label{upper bound of Vapp}
\Vapp(\theta)  \leq \frac{1}{\gamma},
\end{equation} 
where $\gamma$ is a user-defined positive constant. Every parameter vector $\theta$ for which the performance degradation is less than $1/\gamma$ can be considered as an acceptable parameter from an application's point of view. Therefore, we define the set of all acceptable parameters, the \emph{application set} as
\begin{equation}
\label{application set}
\Theta(\gamma) = \left\{ \theta | \Vapp(\theta)  \leq \frac{1}{\gamma} \right\}.
\end{equation}
The application set (\ref{application set}) has been extensively used in applications oriented input design for system identification (see \cite{BoWahlbergHakanHjalmarson2010}, \cite{Larsson2011a} and \cite{Hjalmarsson:2009}). The main objective of applications oriented input design is to provide a tool for designing the input signal to be used in the identification experiment such that the estimated model guarantees acceptable control performance when used in the control design, that is, we want $\hat{\theta}_N\in\Theta(\gamma)$ with high probability. This requirement can be formulated mathematically as the set constraint
\begin{equation}
\label{id in app}
\Esi(\alpha) \subseteq \Theta(\gamma). 
\end{equation}
Therefore, the input design problem can be formulated as an optimization problem, where (\ref{id in app}) plays the role of a constraint. However, one crucial issue is that while $ \Esi$ is an ellipsoidal set, the application set can be of any shape. Thus, the set constraint (\ref{id in app}) may not be convex. Two known approaches to make a convex approximation of the constraint are discussed in the next section. Alternatives to constraint (\ref{id in app}) can be found in \cite{Rojas2011}.
\section{Application set approximation}
\label{Application set approximation}
Two methods of approximating the set constraint with a convex one are the scenario approach, see \cite{Larsson2011}, \cite{Calafiore2006}, and the ellipsoidal approach, see \cite{BoWahlbergHakanHjalmarson2010}.

In the scenario approach, the application set is described by a number, $N_k$, of samples (or scenarios) which are randomly chosen from the set. The constraint (\ref{id in app}) is then replaced by a set of inequalities,
\begin{equation}
[\theta-\theta_0]^T I_F(\theta_0)[\theta-\theta_0] \hspace{-0.1cm}\geq\hspace{-0.1cm}  \frac{\gamma \chi^2_\alpha(n)}{N} V_{app}(\theta_k), k = 1, \ldots, N_k.
\end{equation}
However, in order to have a good approximation of the application set, the number of samples must be large enough (see e.g. \cite{Campi2008} for the minimum required number of scenarios). This is not easy to satisfy, especially in high dimensional and complex plants, since for certain controllers, such as MPC, it is not possible to find analytic expressions for $\Vapp$. Therefore, a large number of highly time-consuming and costly simulations are necessary.

The ellipsoidal approach is based on a second order Taylor expansion of $\Vapp(\theta)$ around $\theta_0$, that is,
\begin{equation}
\label{ellipsoidal approx}
\begin{split}
\Vapp(\theta) &\approx \Vapp(\thetao) + \Vapp'(\theta) [\theta - \thetao] 
\\
& +  0.5 [\theta - \thetao]^T\Vapp''(\theta) [\theta - \thetao] 
\\
 &= 0 + 0 + 0.5 [\theta - \thetao]^T \Vapp''(\theta) [\theta - \thetao] .
\end{split}
\end{equation}
The application set can thus be approximated by the ellipsoidal set
\begin{equation}
\label{2nd taylor}
\Eapp (\gamma) = \left\{\theta | [\theta - \thetao]^T V^{''}_{app}(\theta) [\theta - \thetao]  \leq \frac{2}{\gamma} \right\}.
\end{equation}
The quality of the approximation not only depends on the application cost but also on the value of $\gamma$. For sufficiently large values of $\gamma$, $\Eapp$ gives an acceptable approximation while for smaller values, higher order terms of Taylor expansion may need to be considered \cite{Hjalmarsson2011}. However, calculation of the Hessian matrix is a challenging task. In many problems it is not possible to analytically determine the Hessian of the application function due to nonlinearities in the controllers that are being used. Therefore, numerical approximations are used. Using numerical methods, such as finite difference approximation, is not possible in many cases because of the large number of variables involved.

\section{Application set approximation for MPC}
\label{proposed method}

MPC has drawn much attention in control fields, thanks to its ability to cope with system limitations. Using MPC, we can deal with both input and output constraints explicitly during the controller design and implementation. However, the resulting explicit solutions for MPC are difficult to deal with due to these constraints, which makes it unavoidable to use numerical calculations for the approximation of the application cost \cite{Larsson2011a}. In this section we present a new approach based on analytical methods for the application cost approximation for MPC. The proposed approach leads to faster estimations of the application sets.

\subsection{Application Cost Function for MPC}

\begin{figure}[thpb]
      \centering
     \includegraphics[scale = 0.25]{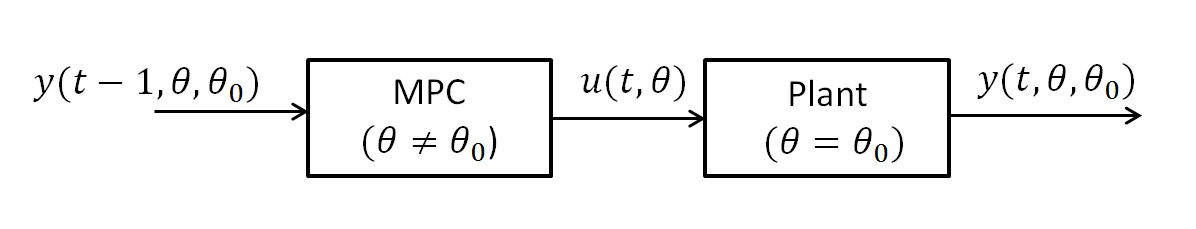}
      \caption{The output signal used in the application cost function (\ref{Vapp for MPC}). }
      \label{Vapplication}
\end{figure}

The application cost function measures the amount of performance degradation that stems from plant-model mismatch. One reasonable choice of this function for MPC is the difference between the measured output when the controller is working based on the true parameters, $\theta_0$, and when it is using perturbed parameters $\theta$, that is,
\begin{equation}
\label{Vapp for MPC}
\Vapp(\theta) = \frac{1}{M} \sum_{t=1}^M \| y(t,\thetao,\thetao)-y(t,\theta,\thetao)  \|^2,
\end{equation}
where $M$ is the number of measurements used, $t$ is time, the second argument of $y$ describes the parameters which are used by MPC and the third one represents the true system parameters \cite{Larsson2011a}. This is shown in Fig. \ref{Vapplication}. However, in reality the true system parameters are not known. Moreover, it is not possible to run the process based on perturbed parameters and measure the real output, since the plant is then controlled using an arbitrary model and it may damage it. Therefore, the following approximation of the application cost is used \cite{Larsson2011a}
\begin{equation}
\label{Vapp est for MPC}
\widehat{V}_{app}(\theta) = \frac{1}{M} \sum_{t=1}^N \| y(t,\hat{\theta},\hat{\theta})-y(t,\theta,\hat{\theta})  \|^2,
\end{equation}
where $\hat{\theta}$ is the best available estimation of $\theta_0$ in the linear approximation of the true system. Thus, the evaluation is done in simulations instead of the real plant.
In addition, the value of $\gamma$ in (\ref{upper bound of Vapp}) is computed based on the idea developed in \cite{Larsson2011a} and we explain it later in the result section.

\subsection{Application Function Approximation}
In order to obtain a convex approximation of the application set, we start by estimating $y(t,\theta,\hat{\theta})$ in (\ref{Vapp est for MPC}). Using a Taylor expansion of $y(t,\theta,\hat{\theta})$, we can write
\begin{equation}
\begin{split}
\label{taylor}
y(\theta) = y(\hat{\theta}) &+ \sum_{i=1}^n \frac{\partial y(\hat{\theta})}{\partial \theta_i}\delta\theta_i 
\\
&+ \frac{1}{2}\sum_{i=1}^n \sum_{j=1}^n \frac{\partial^2y(\hat{\theta})}{\partial \theta_i \partial \theta_j}\delta\theta_i \delta\theta_j+ hot,
\end{split}
\end{equation}
where $\theta_i$ are the elements of $\theta$. Here, the first and third arguments of $y(t, \theta, \hat{\theta})$, are omitted for the sake of simplicity. In order to find the derivatives in (\ref{taylor}), we need to find the derivatives of the input signal generated by MPC with respect to $\theta$. However, this is a challenging problem since the solution of MPC is not simple enough when there are inequality constraints on input and output signals. The proposed solution here is to notice that, when $\theta$ is a small perturbation of $\hat \theta$, the active constraints are the same as when the MPC is based on $\hat \theta$. Thus, the main idea is to let MPC run based on $\hat{\theta}$ at each time instance $t$, and determine the optimal value of the input signal $u(t,\hat{\theta})$.
We assume that the active constraints remain the same for small perturbations of $\hat{\theta}$. Therefore, at time step $t$, we are able to find an explicit solution of the optimization problem in MPC for $\theta = \hat{\theta} + \delta \theta$ by considering active constraints as equality constraints. We can analyze the effects of perturbing the parameters when $\delta \theta$ is small enough.
In the rest of this section, we briefly describe the explicit solution of MPC when we are considering only active constraints, then we provide insights into the perturbation analysis for the MPC solution. Finally, we show how these concepts can be used to find the derivatives in (\ref{taylor}) and compute the application cost function.

\subsubsection{Explicit Solution of MPC}
Consider the MPC problem (\ref{MPCOID}) at time instance $t$. For simplicity it is assumed that $N_u = N_y$. Now we seek to rewrite the MPC formulation as a quadratic program where we are considering only active constraints obtained by solving MPC for $\hat{\theta}$, which are equality constraints. Introducing
\begin{eqnarray}
\begin{aligned}
&X(\theta)  =
[x(N_u+1,\theta)^T, \ldots , x(1,\theta)^T,u(N_u,\theta) , \ldots , u(1,\theta)^T]^T,
\nonumber\\
&\Delta =
\begin{bmatrix}
      I        &       -I        &    \cdot      &    0    \\
  \vdots    &     \ddots    &    \ddots    & \vdots\\
     0        &     \cdot      &        I        &    -I     \\
     0        &     \cdot      &        0       &     I     \\
\end{bmatrix},
\nonumber
\Upsilon =
\begin{bmatrix}
I_{N_u+1} \otimes C(\theta)         &&                     0                \\
                    0                             &&    I_{N_u} \otimes \Delta  \\
\end{bmatrix},
\nonumber\\
&\mathcal{Q} =
\begin{bmatrix}
I_{N_u+1} \otimes Q     &&             0              \\
                 0                 &&   I_{N_u} \otimes R\\
\end{bmatrix},
\nonumber\\
 &\mathcal{H}=
[r(N_u+1)^T , \ldots , r(1)^T , 0 , \ldots , 0, u^{*}(t-1,\theta)^T]^T,
\end{aligned}
\end{eqnarray}
where by $I_m \otimes M$, we mean the Kronecker products of $I_m$ and $M$ \cite{Laub2005}, we can rewrite the cost function $J$ in (\ref{MPCOID}) in the following form:
\begin{equation}
J = (\Upsilon (\theta) X(\theta) - \mathcal{H})\mathcal{Q}(\Upsilon (\theta) X(\theta) - \mathcal{H})^{T}.
\end{equation}
Moreover, the system dynamics and the first equality constraint in (\ref{MPCOID}), give that $\mathcal{C}(\theta)X(\theta) = \mathcal{D}(\theta)$, with
\begin{equation}
\begin{split}
\mathcal{C} \hspace{-0.1cm}&=
\begin{bmatrix}
I         \hspace{-0.3cm}& -A(\theta)     \hspace{-0.3cm}&  \ldots  \hspace{-0.3cm}&     0        \hspace{-0.3cm}&       0       \hspace{-0.3cm}&   -B(\theta) \hspace{-0.3cm}&   \ldots     \hspace{-0.3cm}&       0       \\
\vdots \hspace{-0.3cm}& \vdots          \hspace{-0.3cm}& \ddots  \hspace{-0.3cm}&   \vdots   \hspace{-0.3cm}&   \vdots     \hspace{-0.3cm}&   \vdots     \hspace{-0.3cm}&    \ddots   \hspace{-0.3cm}&     \vdots   \\
0        \hspace{-0.3cm}&    0              \hspace{-0.3cm}& \ldots  \hspace{-0.3cm}&        I      \hspace{-0.3cm}&  -A(\theta) \hspace{-0.3cm}&        0       \hspace{-0.3cm}&    \ldots     \hspace{-0.3cm}&  -B(\theta) \\
0        \hspace{-0.3cm}&    0              \hspace{-0.3cm}& \ldots  \hspace{-0.3cm}&       0      \hspace{-0.3cm}&      I         \hspace{-0.3cm}&        0        \hspace{-0.3cm}&    \ldots    \hspace{-0.3cm}&      0    
\end{bmatrix},
\\
\mathcal{D}\hspace{-0.1cm} &=
\begin{bmatrix}
0 \\
\vdots \\
0 \\
\hat{x}(t,\theta)
\end{bmatrix}.
\end{split}
\end{equation}
Now consider the inequality constraints in (\ref{MPCOID}). They can be rewritten as
\begin{equation}
\label{inequalities}
\begin{bmatrix}
I_{N_u+1} \otimes C(\theta)  &  0\\
-I_{N_u+1} \otimes C(\theta) &  0\\
              0                           &   I\\
              0                           &  -I\\
\end{bmatrix}
X(\theta) \leq
\begin{bmatrix}
I_{N_u+1}  \otimes y_{max} \\
-I_{N_u+1} \otimes y_{min} \\
I_{N_u}      \otimes u_{max} \\
-I_{N_u}    \otimes u_{min} \\
\end{bmatrix}.
\end{equation}
Let $\Xi$ be a diagonal matrix, where each diagonal element corresponds to one of the inequality constraints in (\ref{inequalities}). A diagonal element is zero if its corresponding constraint is inactive and it is one for active constraints. Multiplying (\ref{inequalities}) by $\Xi$ and introducing
\begin{equation}
\begin{split}
\Xi_a &= \Xi
\begin{bmatrix}
I_{N_u+1} \otimes C(\theta)  &  0\\
-I_{N_u+1} \otimes C(\theta) &  0\\
              0                           &   I\\
              0                           &  -I\\
\end{bmatrix},\\
\nonumber
\rho &= \Xi
\begin{bmatrix}
I_{N_u+1}  \otimes y_{max} \\
-I_{N_u+1} \otimes y_{min} \\
I_{N_u}      \otimes u_{max} \\
-I_{N_u}    \otimes u_{min} \\
\end{bmatrix},
\end{split}
\end{equation}
we get $\Xi_a = \rho$, which represents those inequality constraints that are active at time instance $t$. Then we can rewrite the entire set of constraints as $\mathcal{A}(\theta)X(\theta) = \mathcal{B}(\theta)$, where\\
$$\mathcal{A}(\theta) = 
\begin{bmatrix} \mathcal{C}(\theta)\\
\Xi_a
\end{bmatrix} , \hspace{0.1cm}
\mathcal{B}(\theta) = 
\begin{bmatrix} \mathcal{D}(\theta)\\
\rho
\end{bmatrix} .
$$
Finally, the following optimization problem is obtained:
\begin{equation}
\begin{split}
\label{MPC_new}
\displaystyle \min_{\substack{X(\theta)}} \hspace{0.3cm}  &(\Upsilon (\theta) X(\theta) - \mathcal{H})\mathcal{Q}(\Upsilon (\theta) X(\theta) - \mathcal{H})^{T},
\\
\text{s.t. } \hspace{0.3cm}
&\mathcal{A}(\theta)X(\theta) = \mathcal{B}(\theta).
\end{split}
\end{equation}
Problem (\ref{MPC_new}) is a quadratic optimization problem with equality constraints. 
The KKT conditions \cite{Boyd2012} for this problem are
\begin{equation}
\begin{split}
& 2\Upsilon(\theta)^T \mathcal{Q} (\Upsilon (\theta) X(\theta) - \mathcal{H}) + \mathcal{A}^T(\theta)\lambda = 0 ,
\nonumber\\
& \mathcal{A}(\theta)X(\theta) = \mathcal{B}(\theta),
\end{split}
\end{equation}
where $\lambda$ are the Lagrange multipliers. This can be written as
\begin{equation}
\label{KKT}
\hspace{0cm}
\begin{bmatrix}
2\Upsilon(\theta)^T \mathcal{Q} \Upsilon (\theta)  \hspace{-0.2cm}&   \mathcal{A}(\theta)^T\\
\mathcal{A}(\theta)                                             \hspace{-0.2cm}&               0
\end{bmatrix}\hspace{-0.15cm}
\begin{bmatrix}
X(\theta)\\
\lambda
\end{bmatrix}
\hspace{-0.12cm}=\hspace{-0.12cm}
\begin{bmatrix}
2\Upsilon(\theta)^T \mathcal{Q} \mathcal{H} (\theta)\\
\mathcal{B}(\theta)
\end{bmatrix},\hspace{-0.2cm}
\end{equation}
or equivalently
\begin{equation}
\label{closedform}
\Psi(\theta) 
\begin{bmatrix}
X(\theta)\\
\lambda
\end{bmatrix}
=
\Lambda(\theta).
\end{equation}
Since the block matrices in $\Psi(\theta)$ are not invertible, (\ref{closedform}) can be solved using the pseudoinverse and Schur complement of the resulting block matrix
\begin{equation}
\label{pinv}
\begin{bmatrix}
X(\theta)\\
\lambda
\end{bmatrix}
=
(\Psi(\theta)^T\Psi(\theta))^{-1}\Psi(\theta)^T\Lambda(\theta).
\end{equation}
Solving (\ref{pinv}), we can easily obtain an explicit solution, $X(\theta)$, for (\ref{MPC_new}).

\subsubsection{Perturbation Analysis}
The analysis in this section are based on the perturbation analysis techniques in \cite{Cao2009} and \cite{Y.C.HO1979}. Having the MPC solution at time step $t$ as a function of $\theta$, our aim is to compute the derivatives of $X(\theta)$ with respect to $\theta$, based on which the derivatives in (\ref{taylor}) will be calculated. This can be obtained by linearizing $X(\theta)$ around $\hat{\theta}$, invoking the Taylor expansion
\begin{equation}
\begin{split}
\label{taylor X}
X(\theta) = X(\hat{\theta}) &+ \sum_{i=1}^n \frac{\partial X(\hat{\theta})}{\partial \theta_i}\delta\theta_i 
\\
&+ \frac{1}{2}\sum_{i=1}^n \sum_{j=1}^n \frac{\partial^2X(\hat{\theta})}{\partial \theta_i \partial \theta_j}\delta\theta_i \delta\theta_j+ hot \hspace{0.1cm},
\end{split}
\end{equation}
where $\theta=\hat\theta+\delta \theta$. Moreover, the Taylor expansion of $X(\theta)$ can be computed writing the Taylor expansions of $\mathcal{A}(\theta)$, $\mathcal{B}(\theta)$, $\Upsilon(\theta)$, and $\mathcal{H}(\theta)$, which in turn are easily derived by having the derivatives of $A(\theta)$, $B(\theta)$, $C(\theta)$, $\hat{x}(t,\theta)$, and $u^*(t-1,\theta)$. The derivatives of $\hat{x}(t,\theta)$, and $u^*(t-1,\theta)$ are available from the Taylor expansion of $X(\theta)$ in the previous time instances.
Now, recall the definition of $y(t,\theta,\hat{\theta})$ and the linear model used for description of the plant
\begin{equation}
\begin{split}
\label{plant output}
x(t+1,\theta)&=A(\hat{\theta})x(t,\theta)+B(\hat{\theta})u(t,\theta),\\
y(t,\theta,\hat{\theta})&=C(\hat{\theta})x(t,\theta),
\end{split}
\end{equation}
where $u(t,\theta)$ is the optimal input designed by MPC. We aim to find the coefficients in (\ref{taylor}). They can be calculated easily in a recursive manner by differentiating (\ref{plant output}) with respect to $\theta$, using the derivatives of $u(t,\theta)$, which are available from (\ref{taylor X}).

\subsubsection{Application Cost Function}
Recall the application function (\ref{Vapp est for MPC}), we can calculate the Hessian matrix in terms of the obtained derivatives of $y$ as follows
\begin{equation}
\label{1st approx}
\begin{split}
\widehat{V}^{''}_{app}(\theta) &= \frac{2}{M} \sum_{t=1}^M \{\frac{\partial y(t, \hat{\theta})}{\partial \theta}\}^T \{\frac{\partial y(t, \hat{\theta})}{\partial \theta}\}\\
& + \frac{2}{M} \sum_{t=1}^M \{\frac{\partial ^2 y(t, \hat{\theta})}{\partial \theta^2}\}^T \{y(t,\hat{\theta},\hat{\theta})-y(t,\hat{\theta},\hat{\theta}) \}.
\end{split}
\end{equation}
Note that the second term is zero since $V_{app}(\hat{\theta}) = 0$. Substituting (\ref{1st approx}) into (\ref{2nd taylor}), we a convex approximation of the application set.

The method provides a fast tool for convex approximation of application cost function. Many calculations in different time instants are the same and can be pre-computed. Moreover, the active constraints may not change often, thus, at each time instance a large number of the calculations can be skipped by re-using the results from previous time instances. Therefore, the proposed approach is much faster than both the scenario-based approach and the ellipsoidal approximation method.

\section{Numerical examples}
\label{Numerical examples}
In this section we evaluate the proposed method in Section~\ref{proposed method} with two numerical examples. 
\subsection{Example 1}
Consider the following system:
\begin{equation}
\begin{split}
x(t+1) &= \theta_2 x(t) + u(t),\\
y(t) &= \theta_1 x(t).
\end{split}
\end{equation}
The true system is given by the parameter values $\theta_0 = [0.6 \hspace{0.2cm} 0.9]^T$. The objective is to find the application set $\Theta$, when MPC is used for reference tracking. We use the MPC formulation in (\ref{MPCOID}), with the following settings:
$N_u = N_y = 5$, $Q = 10$, $R = 1$, $u_{max}=-u_{min}=1$, $y_{max}=-y_{min}=2$.

\begin{figure}[thpb]
      \centering
     \includegraphics[scale = 0.65]{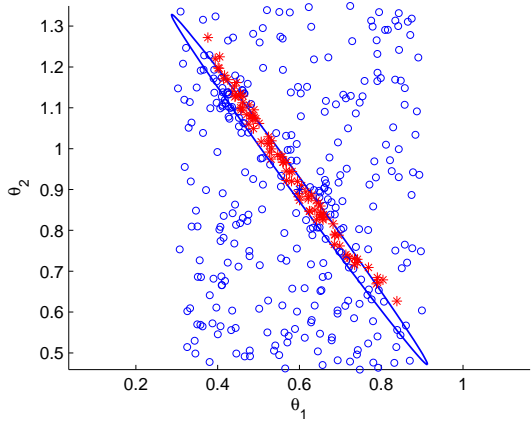}
      \caption{Approximated $\epsilon_{app}$ ('\textcolor[rgb]{0.00,0.00,1.00}{-}') and 400 randomly generated samples of $\theta$. '\textcolor[rgb]{1.00,0.00,0.00}{$\ast$}' represents samples which are satisfying $V_{app}(\theta) \leq \frac{1}{\gamma}$ while '$o$' are those located outside the application set. 81\% of the samples that fulfill the condition $V_{app}(\theta) \leq \frac{1}{\gamma}$ are located inside the approximated ellipsoid given by the proposed method.}
      \label{theta}
\end{figure}

We set the length of the experiment to $N = 100$ samples and the accuracy to $\gamma = 1000$. Note that we use the application cost function defined in (\ref{Vapp for MPC}). Now using the proposed approach, we obtain the application ellipsoid shown in Fig. \ref{theta}.

In order to check the accuracy of the proposed method, we perform 400 simulations with different values of $\theta$ which are generated randomly with a uniform distribution. The results show that from 400 generated points, 93 points are satisfying the condition $V_{app}(\theta) < \frac{1}{\gamma}$. Among all accepted values of $\theta$, $81\%$ are completely inside or on the border of the approximated ellipsoid, which means that the estimated ellipsoid covers at least 81\% of the acceptable.

Furthermore, the Hessian matrix is computed employing numerical methods, provided by DERIVESTsuite. The application set is then approximated using the ellipsoidal approach (\ref{2nd taylor}). As expected, the result is the same as when the proposed method is used. However, in the proposed method, we need only one complete simulation of the closed loop system with MPC, while in the numerical approximation of the Hessian, which is based on finite difference approximation, $O(6*n^2)$ number of simulation is required depending on the selected accuracy. Therefore, the new approach is expected to be faster. While it takes 94 seconds for the numerical method to calculate the Hessian matrix in this example, the new method needs only 12 second to give the same approximation, which means that 87\% of time is saved.

\subsection{Example 2}
In this example we illustrate the algorithm on a more complex and experimental example. We consider a distillation column. The nonlinear system representation is taken form a benchmark process proposed by the Autoprofit project \cite{Autoprofit} is used. For a general description of distillation columns, we refer the reader to \cite{SkogestadDC}.

The plant is linearized around the steady state operating conditions and then, using model order reduction methods, the second order model
\begin{equation}
\begin{split}
x(t+1) &= 
\begin{bmatrix}
\theta_1 & \theta_2\\
\theta_3 & \theta_4
\end{bmatrix}
 x(t) + 
\begin{bmatrix}
\theta_5 & \theta_6\\
\theta_7 & \theta_8
\end{bmatrix}
u(t),\\
y(t) &= 
\begin{bmatrix}
-0.8954  & 0.1421\\
-0.2118 & -0.1360
\end{bmatrix}
 x(t) + e(t),
\end{split}
\end{equation}
is obtained, where, $e(t)$ is a white measurement noise with variance $E\{e(t)^Te(t)\}=0.001$.
We assume that $1\%$ performance degradation from the case when MPC is using the true parameters is allowed, that is,
\begin{equation}
\gamma = \frac{100}{V(\theta_0)},
\nonumber
\end{equation}
where
$V(\theta_0) = \frac{1}{M}\sum_{t=1}^{M}\|y(t,\theta_0,\theta_0) - r(t)\|^2$, see \cite{Larsson2011a}.

\begin{figure}[thpb]
      \centering
     \includegraphics[scale=0.6]{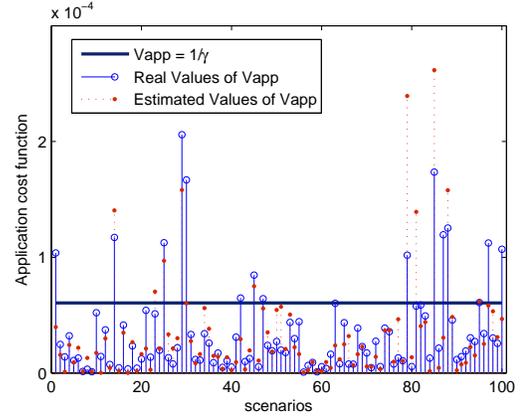}
      \caption{Approximated  ('\textcolor[rgb]{0.00,0.00,1.00}{o --}') and real ('\textcolor[rgb]{1.00,0.00,0.00}{. - -}') values  of $V_{app}(\theta)$ for 100 different samples of $\theta$ taken form a uniform distribution.}
      \label{Vapp_all}
\end{figure}

\begin{figure}[thpb]
      \centering
     \includegraphics[scale=0.6]{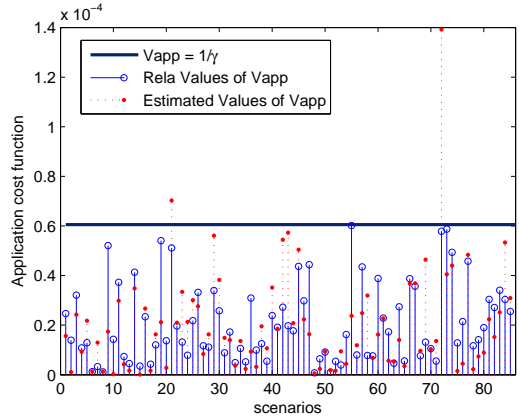}
      \caption{Approximated  ('\textcolor[rgb]{0.00,0.00,1.00}{o --}') and real ('\textcolor[rgb]{1.00,0.00,0.00}{. - -}') values  of $V_{app}(\theta)$ inside the application set. 92\% of the samples inside the region are classified as acceptable ones by the proposed method.}
      \label{Vapp_in}
\end{figure}

Since MPC is used for tracking, the model is augmented with a constant output disturbance on each output to get integral action. This is presented in further detail in \cite{J.M.Maciejowski2002}. 
The proposed method has been employed to calculate the approximate application cost in (\ref{ellipsoidal approx}). In order to evaluate the capability of the method, we run the process for 100 different values of $\theta$, taken from a uniform distribution. Fig. \ref{Vapp_all} shows the real and approximated values of the application cost function for each scenario. 
In order to have a better insight, the samples which are located inside the application set are illustrated in Fig. \ref{Vapp_in}. It can be easily seen that the proposed method has a good performance inside the application set. Among 85 scenarios which result in an acceptable application cost, 83 scenarios are approximated as acceptable ones using the proposed method. The method classifies 6 points outside the region as acceptable ones. Therefore, the obtained accuracy of the proposed method is 92\%.

\section{Conclusions}
\label{Conclusions}
In this paper we have introduced a general technique for the approximation of the application set, a structure required for the implementation of optimal input design schemes. In particular, we have focused on MPC, a control technique for which it is not possible to obtain the application set explicitly. Some simulation examples have been presented, which show the advantages of the new method with respect to previous techniques,in terms of speed.

The method is general enough to be applied to other controller strategies and application areas where it is not possible to derive the application set explicitly. Specifically, the method can be extended to MPC for nonlinear plants, with more complicated noise structures, and the derivation of expressions for higher order derivatives of the cost function could be used, in principle, to obtain better approximations of the application set using techniques such as the one presented in \cite{Hjalmarsson2011}.

\addtolength{\textheight}{-3cm}   

%
\bibliographystyle	{ieeetransactions_no_urls}		
\bibliography{CDCref,referenslista,refs,refs2,myrefs2}
%

\end{document}